\documentclass[9pt,twocolumn,twoside]{pnas-new-arxiv}
\usepackage{amsmath}
\usepackage{xeCJK}

\setboolean{displaywatermark}{false} 
\fancypagestyle{firststyle}{}
\templatetype{pnasresearcharticle-arxiv} 

\begin{document}

\title{Full-component reconstruction of three-dimensional fluid stress tensors}

\author[a]{Shunsuke Kumagai}
\author[a]{Shun Miyatake}
\author[a]{Ryusuke Cho}
\author[a]{William Kai Alexander Worby}
\author[a]{Masanori Naito}
\author[a]{Takahiro Ushioku} 
\author[b]{Masanobu Horie}
\author[a*]{Yoshiyuki Tagawa}

\affil[a]{Department of Mechanical Systems Engineering, Tokyo University of Agriculture and Technology, Koganei, Tokyo, 184-8588, Japan}
\affil[b]{RICOS Co. Ltd., Chiyoda-ku, Tokyo 1000005, Japan}

\leadauthor{Kumagai}


\authorcontributions{Y.T. conceived and supervised the study. Y.T. and M.H. designed the research. S.K., S.M., R.C., and W.K.A.W. built the experimental system and performed the experiments. S.K., M.N., T.U., and M.H. developed the machine-learning methodology. W.K.A.W. contributed to the theoretical framework of photoelastic stress measurement. S.K. analyzed the data. S.K. and Y.T. wrote the paper with input from W.K.A.W. and T.U. All authors discussed the results and edited the manuscript. Y.T. acquired funding.}
\authordeclaration{The authors declare no competing interests.}
\acknow{This work was supported by JSPS KAKENHI Grant No. JP24H00289 (Y.T.) and JST PRESTO Grant No. JPMJPR21O5 (Y.T.), Japan.}
\correspondingauthor{\textsuperscript{*}To whom correspondence should be addressed. E-mail: tagawayo@cc.tuat.ac.jp \color{black}}


\begin{abstract}
Forces govern how fluids deform biological tissues, regulate cardiovascular function, and determine the performance and failure of soft materials.
Recent advances in flow birefringence, including the use of suspended anisotropic nanomaterials to optically encode stress in fluids, have made direct stress measurement experimentally accessible in projection.
However, direct experimental access to all six components of the three-dimensional (3D) fluid stress tensor has remained unattainable because optical measurements provide only path-integrated observables.
Recovering local 3D stresses from such data constitutes an intrinsically underdetermined tensor tomography problem, in which two optical observables must determine six independent stress components.
Here we introduce U-FlowPET, an unsupervised physics-informed framework that integrates photoelastic tomography with the governing equations of fluid mechanics to reconstruct the full 3D stress tensor without relying on constitutive assumptions, geometric symmetry, or labeled training data.
Rather than learning from labeled reference stress fields, the method identifies physically admissible stress fields that satisfy momentum balance and continuity while remaining consistent with measured optical projections.
We validate the approach using analytical, numerical, and experimental datasets. 
In axisymmetric pipe flow with an analytical solution, all six stress components are reconstructed with normalized mean absolute errors below 4\%. 
Robust reconstruction is further demonstrated in geometrically complex curved-pipe flow without symmetry assumptions and in experimental pipe-flow data despite measurement noise.
By enabling direct 3D stress-field reconstruction from optical data alone, U-FlowPET extends fluid analysis from observing motion to quantifying force and establishes a new framework for stress-based diagnostics in biological flows and functional materials.
\end{abstract}


\maketitle
\thispagestyle{firststyle}
\ifthenelse{\boolean{shortarticle}}{\ifthenelse{\boolean{singlecolumn}}{\abscontentformzatted}{\abscontent}}{}

\dropcap{F}or over a century, fluid mechanics has relied primarily on velocity fields to understand flow-driven phenomena. From particle image velocimetry (PIV) \cite{Adrian_1991,Adrian2005,Scarano_2013,Scharnowski2020,Schroder2023} to modern three-dimensional diagnostics, the motion of the tracer in a fluid flow has been measured to infer deformation and transport. 
However, velocity is a kinematic quantity---the result of forces, not their cause. In biological systems, vascular remodeling and cellular mechanotransduction respond to stress rather than velocity \cite{Davies_1995,Davies_Peter_2008,figueroa2009computational,marsden2009evaluation,wilson2005predicting,taylor2009patient,taylor1999predictive}, and in soft robotics and functional materials, mechanical performance is governed by internal force transmission \cite{Blaeser_2016,Xu2022,marchese2014autonomous,marchese2015dynamics,rus2015design}. Direct experimental access to the full three-dimensional (3D) stress tensor field in fluids would therefore mark a transition from kinematic observation to dynamic understanding.

Despite its importance, obtaining all six independent components of the 3D fluid stress tensor remains experimentally challenging. 
The Cauchy stress tensor in three dimensions is represented by a $3\times3$ matrix (nine components), which reduces to six independent components due to symmetry. 
Conventional pressure sensors \cite{liu2025pressure,bucinskas2025integrated} provide only localized wall measurements and cannot resolve internal stress structures non-invasively. 
Velocity-based methods indirectly infer stresses: The strain rate tensor is computed by spatial differentiation of measured velocities, and stresses are estimated through a constitutive relation linking strain rate to stress. 
This procedure amplifies measurement noise and becomes unreliable in flows with unknown or complex constitutive behavior, including non-Newtonian fluids.

Photoelasticity \cite{mcafee1974scattered,aben1997photoelastic,aben2005photoelastic,errapart2007technology,H_T_Jessop_1949,doyle1982nonlinearity} offers a fundamentally different approach by directly encoding stress-induced anisotropy into optical birefringence without requiring spatial differentiation. 
However, the measured photoelastic parameters—retardation $\Delta$ and principal orientation $\phi$—are path-integrated projections of stress along the optical axis, and recovering local 3D stresses from these projections requires tomography. 
Unlike scalar tomography, which reconstructs quantities such as density \cite{schmidt2025twenty,cormack1963representation,gordon1970algebraic,ramachandran1971three}, photoelastic reconstruction constitutes a tensor tomography problem: two optical observables must determine six independent components of a 3D tensor field, rendering the inverse problem intrinsically under-determined.

Historically, this information deficiency confined reconstruction to simplified cases. 
Classical integrated photoelasticity succeeded mainly under axisymmetric or two-dimensional assumptions, where strong geometric symmetry reduces the number of unknowns \cite{aben1982integrated,aben1986integrated,aben1993integrated,doyle1978integrated,chen1990utilizing,lane2024two,li2024dynamic}. 
Extensions to arbitrary geometries proved unstable under limited viewing angles and experimental noise \cite{hammer2004reconstruction,lionheart2009reconstruction,szotten2011limited,abrego2019experimental,aben2004optical,aben2012photoelastic}. 
Analytical inversion collapses without imposing constitutive relations, symmetry, or restrictive regularization. 
As a result, full 3D stress tensor reconstruction in flowing fluids has remained unresolved.

Recent advances in inverse problems and machine learning have enabled recovery of high-dimensional fields from incomplete data \cite{RAISSI2019686,RAISSI2020,Cai_2021,JIN2017}. 
Convolutional neural networks and physics-informed frameworks show promise, yet existing stress-reconstruction approaches rely on supervised learning with ground-truth stress fields that are experimentally unavailable \cite{igarashi2025reconstruction,sergazinov2021machine,tao2022photoelastic}. 
The central question therefore remains: 
Can this information-deficient tensor tomography problem be closed without constitutive models, geometric symmetry, or labeled training data?

\begin{figure*}[t]
\centering
\includegraphics[width=2.0\columnwidth]{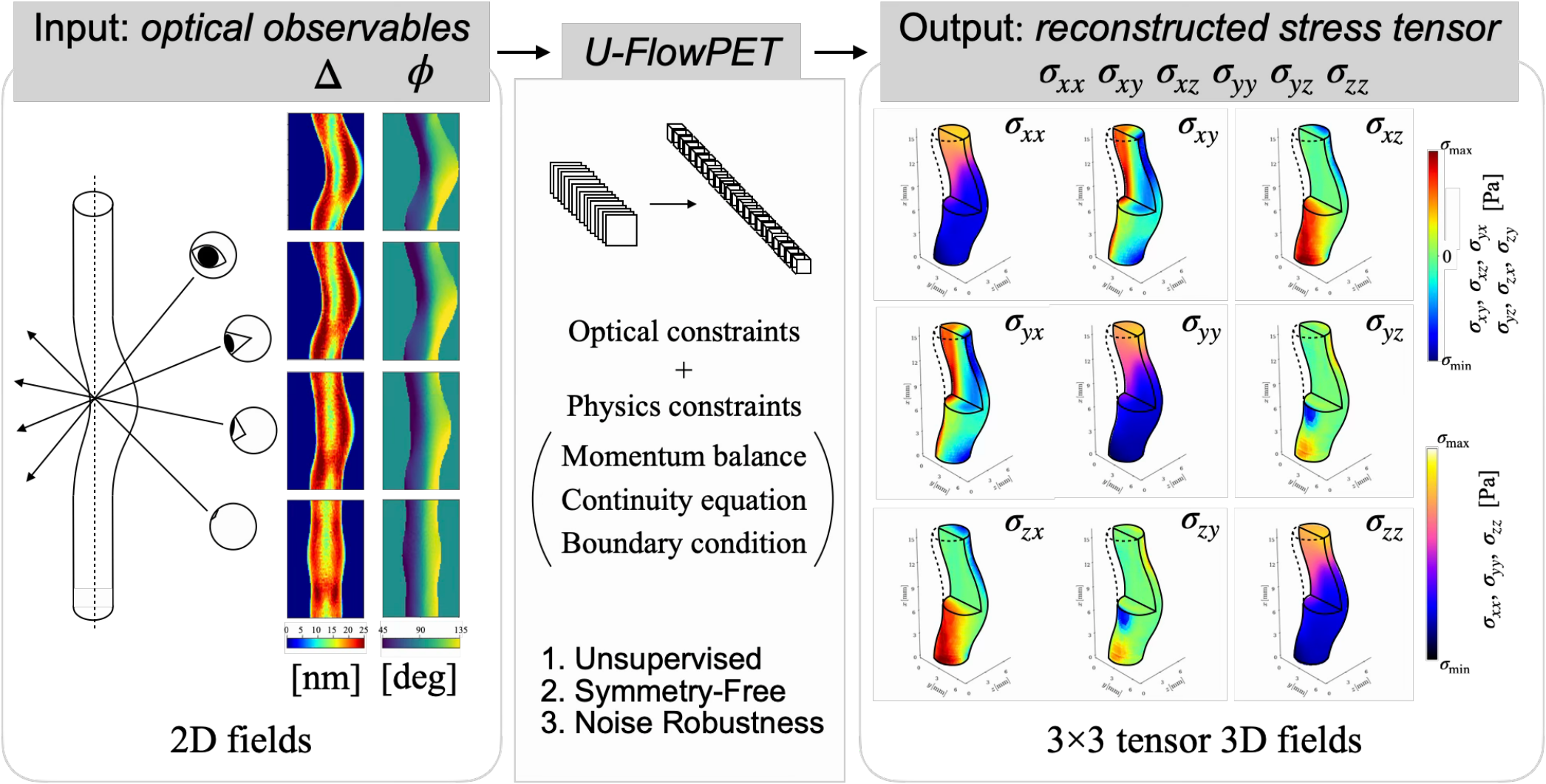}
\caption{Concept of U-FlowPET for full-component reconstruction of three-dimensional fluid stress tensors.
Auxiliary solid and dashed black lines in the fluid stress tensor fields are used to highlight boundaries and cross-sectional contours.
Multi-angle photoelastic measurements provide two optical observables: retardation $\Delta$ and principal orientation $\phi$, as path-integrated projections of stress in a flowing birefringent fluid (\textbf{Movie S1}). 
U-FlowPET integrates optical data consistency with governing fluid mechanics constraints to reconstruct the full three-dimensional Cauchy stress tensor field ($3\times3$ tensor with six independent components). 
The framework operates in an unsupervised manner, without constitutive assumptions, geometric symmetry, or labeled training data, and enables symmetry-free and noise-robust reconstruction of internal stress fields.
}
\label{fig:Overview}
\end{figure*}

Here, we introduce U-FlowPET (Unsupervised flow-integrated photoelastic tomography), a physics-informed unsupervised framework that integrates photoelastic tomography with the governing equations of fluid mechanics to reconstruct  full 3D stress tensor fields. 
Figure \ref{fig:Overview} illustrates the conceptual framework of U-FlowPET. 
Multi-angle measurements of retardation $\Delta$ and principal orientation $\phi$ provide two optical observables that represent path-integrated projections of stress. 
By enforcing consistency with these optical data while simultaneously satisfying momentum balance and the continuity equation of incompressible fluids, the framework reconstructs all six independent components of the three-dimensional Cauchy stress tensor ($\sigma_{xx}, \sigma_{xy}, \sigma_{xz}, \sigma_{yy}, \sigma_{yz}, \sigma_{zz}$) without relying on constitutive relations or geometric symmetry assumptions.
We validate the method via three test cases: synthetic axisymmetric pipe flow with analytical solutions, geometrically complex curved-pipe flows without symmetry assumptions, and experimental pipe flow data in the presence of measurement noise. 
Across these validations, the six stress components are reconstructed with normalized mean absolute errors below 4\% in the analytical pipe-flow benchmark, within 10\%--18\% in the curved-pipe case, and below 8\% for experimentally acquired pipe-flow data, demonstrating that physically constrained unsupervised learning can resolve an intrinsically under-determined tensor tomography problem.
  
By enabling direct 3D stress-field reconstruction without constitutive assumptions or labeled data, U-FlowPET establishes an experimental platform for accessing mechanical evidence in complex flows. 
This framework extends fluid analysis from motion observation to force quantification, opening pathways for stress-based diagnostics in biomedical flows, functional materials, and multiphase transport.

\section*{U-FlowPET}

To close the intrinsically under-determined tensor tomography problem described above, we introduce U-FlowPET (unsupervised flow-integrated photoelastic tomography), a physics-informed unsupervised reconstruction framework that integrates photoelastic tomography with the governing equations of fluid mechanics.

The conceptual structure of U-FlowPET is illustrated in Fig.~\ref{fig:U-FlowPET}. 
Multi-angle measurements of retardation $\Delta$ and principal orientation $\phi$ provide two optical observables that represent path-integrated projections of stress in a flowing birefringent fluid. 
From these projections, U-FlowPET reconstructs the full three-dimensional Cauchy stress tensor field (a $3 \times 3$ tensor with six independent components, namely, $\sigma_{xx}, \sigma_{xy}, \sigma_{xz}, \sigma_{yy}, \sigma_{yz}, \sigma_{zz}$), thereby addressing an intrinsically information-deficient $2 \rightarrow 6$ inverse problem.

A central distinction of U-FlowPET is that stress is treated as an independent field variable rather than being derived from velocity gradients through a constitutive relation. 
In conventional approaches, stresses are obtained from the strain-rate tensor via a constitutive equation linking deformation to stress. 
In contrast, U-FlowPET imposes the Cauchy momentum equation,
\begin{equation}
\rho \frac{D\mathbf{v}}{Dt} =\nabla \cdot \boldsymbol{\sigma} + \mathbf{f},
\label{eq:Cauchy}
\end{equation}
as a physical admissibility constraint, where $\rho$ is the fluid density, 
$\mathbf{v}=(u,v,w)$ is the velocity vector, $\boldsymbol{\sigma}$ is the Cauchy stress tensor, 
and $\mathbf{f}$ is the body force. 
The operator $D/Dt$ denotes the material derivative, defined as $D/Dt = \partial/\partial t + (\mathbf{v}\cdot\nabla)$, which represents the rate of change experienced by a fluid particle moving with the flow.

\begin{figure*}[htbp]
\centering
\includegraphics[width=2.0\columnwidth]{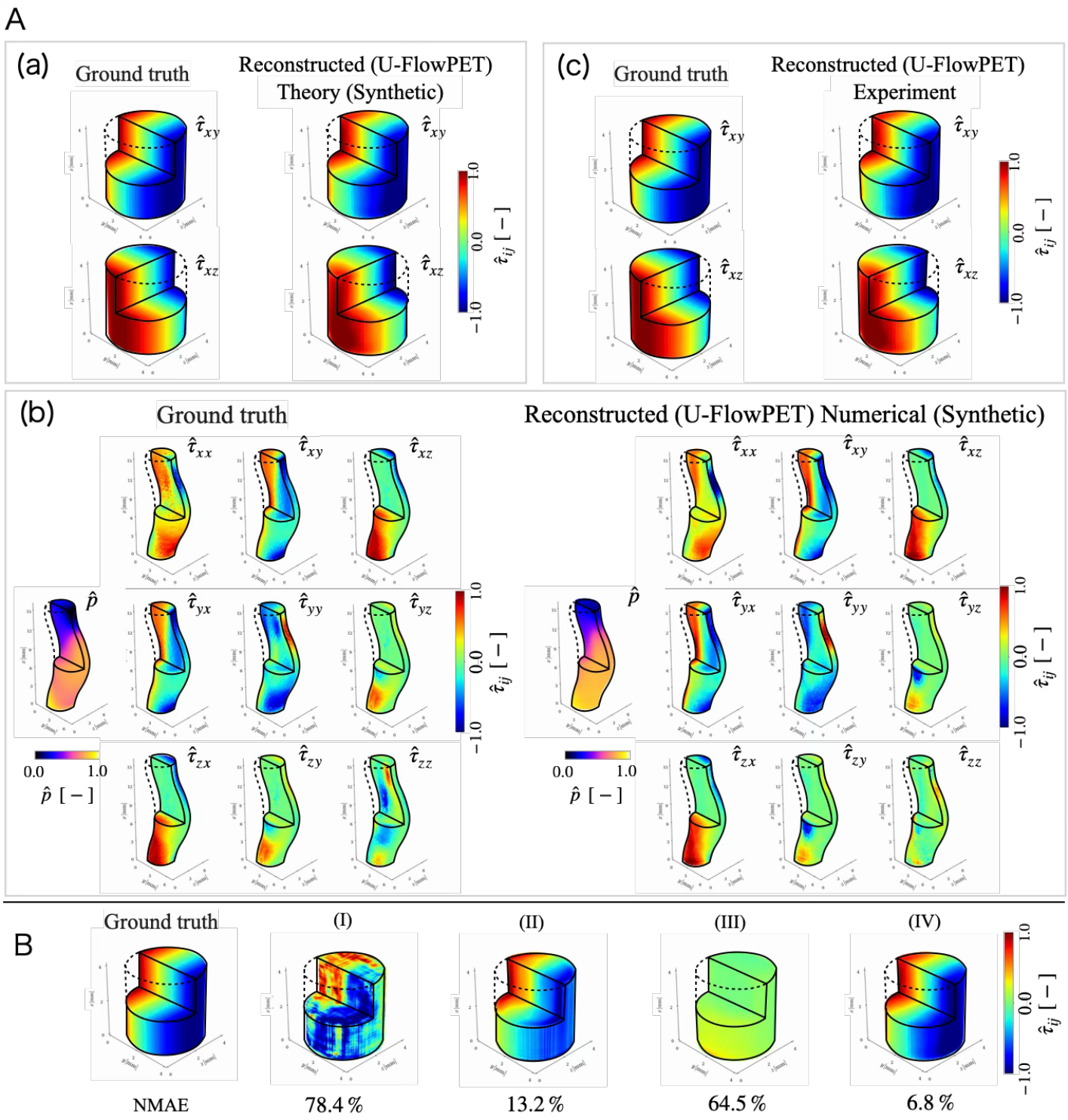}
\caption{Full-component reconstruction of three-dimensional fluid stresses using U-FlowPET.
Auxiliary solid and dashed black lines highlight boundaries and cross-sectional contours.
For visualization, the stress fields are normalized using the minimum and maximum of the corresponding ground-truth fields.
(A) Comparison between ground-truth stress fields and the fields reconstructed by U-FlowPET.
Representative components are shown for three cases:
(a) analytical synthetic pipe flow with a known solution,
(b) numerically generated curved-pipe flow without geometric symmetry, and
(c) experimentally acquired photoelastic data under realistic measurement noise.
The deviatoric stress tensor $\boldsymbol{\tau}$ is mainly visualized here because it represents the flow-induced mechanical structure of interest, whereas the full Cauchy stress tensor $\boldsymbol{\sigma} = -p\mathbf{I} + \boldsymbol{\tau}$ is illustrated in Fig.~\ref{fig:Overview}.
In all cases, the reconstructed fields closely reproduce the spatial structure of the reference distributions.
(B) Ablation study illustrating the roles of optical and physical constraints.
For visualization, the shear component $\tau_{xy}$ is shown, while other components exhibit similar trends.
Reconstructions are compared for four configurations:
(I) retardation only,
(II) retardation and orientation without physical constraints,
(III) physical constraints only, and
(IV) the full U-FlowPET framework.
The optical inputs correspond to experimentally measured birefringence fields such as those shown in Fig.~\ref{fig:Experiments}(B), which contain measurement noise.
Normalized mean absolute errors (NMAE) are reported with respect to a characteristic stress scale defined as the spatially averaged Frobenius norm of the ground-truth deviatoric stress tensor.
The corresponding representative stress scales are 0.30 Pa and 0.32 Pa for the pipe and curved-pipe flows, respectively.
The comparison demonstrates that optical consistency determines stress magnitude and orientation, while physical consistency suppresses nonphysical solutions and noise, enabling robust reconstruction of the stress field.}
\label{fig:results}
\end{figure*}

The Cauchy stress tensor is decomposed as $\boldsymbol{\sigma} = -p\mathbf{I} + \boldsymbol{\tau}$,
where $p$ denotes the isotropic pressure and $\boldsymbol{\tau}$ the extra (deviatoric) stress tensor. 
Figure~\ref{fig:Overview} illustrates the full stress tensor $\boldsymbol{\sigma}$, whereas Fig.~\ref{fig:results} visualizes mainly $\boldsymbol{\tau}$ because the deviatoric stress captures 
the flow-induced mechanical structure of interest.

Importantly, in this formulation, the governing equation does not compute stress from velocity gradients; 
instead, stress is treated as an independent field variable, and the equation acts to eliminate physically 
inconsistent stress fields from the solution space. 
This formulation enables reconstruction without prior knowledge of constitutive models and avoids the error amplification associated with spatial differentiation of velocity measurements.

Operationally, U-FlowPET restricts the admissible solution space through two complementary constraints: 
First, the consistency of the optical projection ensures agreement between the predicted stresses and the measured birefringence. 
Second, physical consistency requires momentum balance and continuity equation of incompressible fluids. 
Together, these constraints select stress fields that are both data-consistent and physically admissible.

The reconstruction network adopts a convolutional encoder–decoder architecture that maps two-dimensional optical patterns to three-dimensional tensor fields.
Hierarchical convolution extracts localized optical features while preserving spatial correlations, which is essential for reconstructing coherent three-dimensional stress structures. 
Similar encoder–decoder strategies have been successfully applied in sparse-view CT reconstruction\cite{bioengineering6040111,Shapira2022ConvolutionalEN} and three-dimensional flow inference\cite{yousif2022deep,Matsuo_2024}; however, unlike supervised approaches, U-FlowPET does not require ground-truth stress labels.

The total loss combines optical and physical consistency terms, enabling unsupervised, symmetry-free, and noise-robust reconstruction. 
Detailed formulations of the forward photoelastic model, governing equations, network architecture, and loss functions are provided in \textit{Materials and Methods}.

\section*{Results \& Discussion}

We evaluated the performance of U-FlowPET in reconstructing three-dimensional stress tensor fields from photoelastic projections. 
Generality and robustness are examined by applying three test cases: analytical synthetic data, geometrically complex synthetic data, and experimental measurements acquired under realistic noise conditions.

\subsection*{Reconstruction from Synthetic Retardation}

U-FlowPET was first applied to two synthetic configurations: an axisymmetric pipe flow with a known analytical solution and a geometrically complex curved-pipe flow without axisymmetry or two-dimensional simplification. 
Figure  \ref{fig:results}(A) compares the reconstructed stress fields with the ground truth.

In the axisymmetric-pipe case, the reconstructed stress components recover the expected spatial distribution of shear and normal stresses along the flow direction. 
For the curved-pipe configuration, despite the absence of geometric symmetry, the model consistently reconstructs spatially varying stress components across the three-dimensional domain. 
In both settings, the reconstructed structures closely match the reference solutions.

Across all six independent stress components, NMAEs, defined with respect to the characteristic stress scale given by the spatially averaged Frobenius norm of the ground-truth tensor, remain below 4\% for the axisymmetric pipe flow and within 10\%--18\% for the curved-pipe case.
These results indicate an accurate recovery of physically significant stress magnitudes and orientations without reliance on ground-truth labels.

The pipe-flow validation demonstrates that U-FlowPET eliminates the need for supervised training data. 
Whereas previous machine-learning approaches required labeled stress fields, reconstruction here emerges solely from optical projection consistency and governing physical constraints.

The curved-pipe case further reveals that geometric symmetry is not required to stabilize the inverse problem. 
Conventional analytical reconstructions reduce the number of unknowns through axisymmetry or two-dimensional assumptions. 
By contrast, U-FlowPET reconstructs the full components of the three-dimensional stress tensor field directly from single-axis rotational scan data.  
Although classical tensor tomography would require multi-axis acquisition to resolve arbitrary geometries, the present framework achieves stable reconstruction under experimentally feasible scanning conditions, substantially reducing measurement complexity.

Residual discrepancies are mainly located near domain boundaries. 
These deviations likely arise from reduced constraint coverage at the boundaries—where stresses are informed from fewer spatial directions—as well as convolutional padding effects in the encoder–decoder architecture. 
Within the interior region, where optical and physical constraints act jointly, the reconstructed fields remain stable and consistent with the ground truth.

Together, these synthetic validations demonstrate that U-FlowPET solves the intrinsically under-determined $2 \rightarrow 6$ tensor tomography problem without constitutive assumptions, symmetry constraints, or labeled training data.

\subsection*{Reconstruction from Experimental Retardation}

We next evaluated U-FlowPET using experimentally acquired photoelastic data obtained under realistic measurement noise. 
In contrast to synthetic datasets, these measurements include optical noise, calibration uncertainty, and finite angular sampling, providing a stringent test of robustness.

The reconstructed three-dimensional stress fields preserve the dominant spatial trends expected for pipe flow, with shear stress distributions consistent with the reference analytical solution.
Despite the presence of measurement noise, the recovered tensor components remain spatially coherent and physically interpretable across the flow domain.

Across all six independent components, NMAEs remain below 8\%, demonstrating that the reconstruction accuracy extends beyond idealized synthetic conditions to experimentally acquired data.

In particular, the physical-consistency term embedded in the loss function acts as a strong regularizer. 
Local fluctuations in the optical measurements are suppressed unless they are supported by the governing equations.
As a result, noise is not merely smoothed numerically but filtered through physical admissibility. 
In contrast, reconstructions based solely on optical projection consistency are highly sensitive to measurement noise and produce unstable stress distributions.

The experimental validation therefore demonstrates that U-FlowPET overcomes a long-standing limitation of photoelastic tensor tomography: sensitivity to measurement noise. 
By enforcing physical constraints, the framework achieves stable and accurate reconstruction from single-axis scan data under realistic experimental conditions.

\subsection*{Ablation Study}

To clarify the origin of the observed robustness to experimental noise, Fig.~\ref{fig:results}(B) shows the results of an ablation study in which the contributions of the optical and physical constraints are isolated.
Four configurations were compared: (I) retardation-only input, (II) retardation and orientation without physical constraints, (III) physical constraints alone, and (IV) the full U-FlowPET framework.

Using retardation alone (I) fails to reconstruct meaningful stress structures, reflecting the severe information deficiency of the inverse problem. 
Incorporating both retardation and principal orientation (II) largely recovers the overall spatial trends; however, the reconstructed fields remain affected by noise amplification near the boundaries and lack full stability.
Conversely, enforcing only the governing fluid equations without optical consistency (III) yields smooth stress distributions, yet the absolute magnitudes and directional features deviate substantially from the ground truth.

These comparisons reveal that the two loss components have complementary roles. 
The optical consistency term determines the magnitude and orientation of stress by anchoring the reconstruction to the measured birefringence data. 
The physical-consistency term, derived from the Cauchy momentum equation and continuity, eliminates physically inadmissible solutions and suppresses high-frequency noise components.

Therefore, robustness against experimental noise does not arise from numerical smoothing alone; rather, it emerges from restricting the reconstruction to the subspace of physically admissible stress fields. These stress fields satisfy the governing equations of motion and therefore correspond to mechanically realizable states that remain consistent with the measured optical projections. 
By selecting solutions that simultaneously satisfy the data consistency and governing equations, U-FlowPET effectively filters noise through physical admissibility, explaining the stability observed in the experimental reconstructions.

\section*{Conclusion}

We have addressed the intrinsically information-deficient tensor tomography problem of estimating six independent stress components from two optical observables by developing U-FlowPET, an unsupervised physics-informed reconstruction framework. 
By integrating photoelastic tomography with the governing equations of fluid mechanics, the proposed method eliminates the need for constitutive relations, geometric symmetry assumptions, and labeled training data that have traditionally constrained stress reconstruction.

Through validation of analytical synthetic flows, geometrically complex numerical configurations, and experimentally acquired data, we demonstrate that U-FlowPET reconstructs all six components of the three-dimensional fluid stress tensor with normalized mean absolute errors below 4\% in the pipe-flow benchmark, within 10\%--18\% in the curved-pipe case, and below 8\% in experimental pipe-flow measurements, without relying on symmetry or supervised labels.
These results establish that full-component stress reconstruction is achievable under experimentally realistic conditions.

The key mechanism underlying this robustness lies in the enforcement of physical admissibility: by constraining the solution space through the Cauchy momentum equation and continuity, physically inconsistent solutions are excluded. 
As a consequence, measurement noise and incomplete data are filtered through physical laws rather than through numerical smoothing alone.

By enabling three-dimensional stress reconstruction directly from optical projections, U-FlowPET extends fluid analysis from observing motion to quantifying force. 
This framework not only provides a foundation for stress-based diagnostics in biological flows, functional soft materials, and complex multiphase systems but also opens new avenues for experimentally accessing mechanical evidence in fluid-driven phenomena.

\matmethods{
This section describes the methodology used to reconstruct the three-dimensional fluid stress tensor field from photoelastic measurements. 
We first summarize the stress–optic law governing birefringence in flowing fluids, then formulate the forward problem of photoelastic tomography, which relates a three-dimensional stress field to two-dimensional optical observables obtained under rotational scanning. 
Next, we describe the experimental photoelastic tomography system and the procedures used to acquire both synthetic and experimental datasets. 
Finally, we present a detailed formulation of U-FlowPET, including the network architecture, loss functions, and physical constraints, and discuss the  limits of applying the present framework.

\subsection*{Stress-Optic Law in Fluids}
\mbox{}

The stress–optic law describes the relationship between fluid stress and the optical birefringence generated by stress-induced molecular alignment in the flow. 
In birefringent fluids, the local stress state modifies the polarization state of transmitted light, producing two measurable optical observables: the phase retardation $\Delta$ and the principal orientation $\phi$ \cite{nakamine2024flow}. 
These quantities encode information about the internal stress field integrated along the optical path [Fig.~\ref{fig:Experiments}(A)].

In the optical system, unpolarized light is first converted into circularly polarized light using a linear polarizer and a quarter-wave plate. 
As this light propagates through the stressed fluid, the local anisotropy induced by stress alters the polarization state. 
After passing through the entire medium, the light becomes elliptically polarized, characterized by a phase retardation $\Delta$ and a principal orientation $\phi$.

The relationship between the internal stress tensor $\boldsymbol{\tau}$ and the measured optical parameters can be expressed through the integrated stress–optic relations. 
Let the material constants $C_1$ and $C_2$ denote the stress–optic coefficients \cite{aben1997photoelastic}.
The two auxiliary quantities $V_1$ and $V_2$ are defined as

\begin{equation}
\begin{split}
    V_1 \equiv \Delta \cos 2\phi
     &= \int \bigg\{ C_1 (\tau_{yy} - \tau_{xx})\\
    &+ C_2 \left[ (\tau_{yy} + \tau_{xx})(\tau_{yy} - \tau_{xx})+ \tau_{yz}^2 - \tau_{xz}^2 \right] \bigg\} dz,
\end{split}
\label{eq:integral_stress_cos} 
\end{equation}
\begin{equation}
\begin{split}
    V_2 \equiv \Delta \sin 2\phi 
    &= \int \bigg\{ 2C_1 \tau_{xy}  \\
    &+C_2 \left[ 2(\tau_{yy} + \tau_{xx})\tau_{xy} + 2\tau_{yz}\tau_{xz} \right] \bigg\} dz,
\end{split}
\label{eq:integral_stress_sin} 
\end{equation}
from which the measurable optical parameters are obtained as

    \begin{equation}
        \Delta = \sqrt{V_1^{2}+V_2^{2}},
        \label{eq:Delta_V1_V2} 
    \end{equation}    

    \begin{equation}
        \phi = \frac{1}{2} \tan^{-1} \frac{V_2}{V_1}.
        \label{eq:phi_V1_V2} 
    \end{equation}  
These relations indicate that the measured optical quantities correspond to path-integrated projections of the internal stress tensor. 
Recovering local 3D stresses from such projections constitutes an intrinsically under-determined tensor tomography problem---an inverse problem in which two observables must determine six independent components of a tensor field.
Such information-deficient tensor tomography problems arise throughout the imaging sciences and remain unresolved in fluid systems.

\subsection*{Computing Phase Retardation from Stress}
\mbox{}

Using the stress–optic law described above, we compute the optical observables—retardation $\Delta$ and principal orientation $\phi$—from a known three-dimensional stress field $\boldsymbol{\tau}$. 
This procedure constitutes the forward model of photoelastic tomography, relating a volumetric stress tensor field to two-dimensional optical measurements obtained under rotational scanning.

Let $\boldsymbol{\tau}$ denote the stress field expressed in the object coordinate system fixed to the flow channel. 
To compute the optical response at a given rotation angle $\theta$, the stress field must be evaluated along the optical path defined in the laboratory frame. 
This requires identifying the position $\mathbf{p}$ inside the rotated flow domain corresponding to each integration point $\mathbf{p'}$ along the optical path and performing the appropriate tensor transformation.

The rotation of the flow domain about the $x$ axis is described by the rotation tensor

    \begin{equation*} \label{eq:affine_transform_detailed}
        \mathbf{R} = \mathbf{R}_x(\theta) = 
            \begin{pmatrix}
                1 & 0 & 0 \\
                0 & \cos\theta & -\sin\theta \\
                0 & \sin\theta & \cos\theta
            \end{pmatrix}.
    \end{equation*}
The coordinates inside the rotated domain are obtained by
    \begin{equation*}
    \mathbf{p} = (\mathbf{p'} - \mathbf{c}) \mathbf{R}^T + \mathbf{c},
    \label{eq:coord_transform_final}
    \end{equation*}
where $\mathbf{c}$ is the rotation center.

Because $\mathbf{p}$ generally does not coincide with a grid point, the stress tensor at $\mathbf{p}$ is evaluated using trilinear interpolation. 
Let $(x_0,y_0,z_0)$ denote the integer part of $\mathbf{p}=(x,y,z)$ and $(x_d,y_d,z_d)$ the fractional part. 
The interpolation weights are

\begin{equation*}
w_{ijk} = (1-x_d)^{1-i} x_d^i (1-y_d)^{1-j} y_d^j (1-z_d)^{1-k} z_d^k.
\label{eq:weight}
\end{equation*}

To prevent interpolation across regions outside the flow domain, a binary mask $M_{ijk}$ is introduced,

\begin{equation*}
M_{ijk} =
\begin{cases}
1 & \text{if } C_{ijk} \neq 0, \\
0 & \text{if } C_{ijk} = 0.
\end{cases}
\label{eq:mask_definition}
\end{equation*}

The interpolated stress tensor is therefore computed as

\begin{equation*}
\boldsymbol{\tau}_{\text{interp}}(\mathbf{p}) =
\frac{\sum_{i,j,k \in \{0,1\}} C_{ijk} w_{ijk} M_{ijk}}
{\sum_{i,j,k \in \{0,1\}} w_{ijk} M_{ijk}} .
\label{eq:masked_interpolation}
\end{equation*}

The stress tensor is then transformed to the laboratory frame according to

\begin{equation*}
\boldsymbol{\tau'} =
\mathbf{R} \, \boldsymbol{\tau}_{\text{interp}}(\mathbf{p}) \, \mathbf{R}^T .
\label{eq:tensor_transform}
\end{equation*}

Substituting $\boldsymbol{\tau'}$ into Eqs.~(\ref{eq:integral_stress_cos})–(\ref{eq:phi_V1_V2}) yields the optical observables $\Delta$ and $\phi$ corresponding to the given rotation angle. 
Repeating this procedure for all rotation angles produces the synthetic optical datasets used for training and validation.

\subsection*{Experimental Setup}
\mbox{}

To acquire the two-dimensional optical observables—retardation $\Delta$ and principal orientation $\phi$—we constructed a photoelastic tomography measurement system [Fig.~\ref{fig:Experiments}(B)]. 
The optical setup consisted of a circularly polarized-light source (wavelength $\lambda=543$ nm), a birefringent flow channel, and a polarization-resolving camera.
A custom-built single-axis
rotation system was used to enable controlled rotation of the liquid-filled channel while maintaining fluid sealing and optical alignment.
Flow actuation, polarization-image acquisition, data storage, and
channel rotation were automated to collect optical projections from
multiple viewing angles reproducibly.

To minimize refraction at the channel wall, the cylindrical flow tube was immersed in an index-matching bath, following the approach reported in Ref. \cite{muto2025development}. 
The refractive index of the surrounding solution was matched to that of the tube material, thereby suppressing optical distortion and enabling accurate measurement of birefringence signals within the flow.

Experiments were conducted in a circular tube with an inner diameter of 4~mm and at a flow rate of $Q=50$~mL/min. 
The working fluid consisted of a suspension of cellulose nanocrystals (0.3 wt.\%) mixed with sodium iodide solution (NaI, 57.6 wt.\%), which exhibits flow birefringence under shear. 
The surrounding index-matching bath consisted of an aqueous NaI solution of the same concentration.

Rheological measurements using a rotational rheometer indicate that the suspension behaves as a power-law fluid with viscosity described by 
$\eta(\dot{\gamma}) = K\dot{\gamma}^{n-1}$, 
with parameters $K = 0.0974$ and $n = 0.2764$. 
Detailed preparation procedures and rheological characterization are provided in the Supplemental Material.

During experiments, the fluid was driven upward through the tube using a syringe pump. 
The transmitted polarization state was recorded by a polarization camera, from which the spatial distributions of retardation and principal orientation were computed. 
The images shown in Fig.~\ref{fig:Experiments}(B) correspond to time-averaged intensity fields under steady flow conditions. 
The operating principle of the polarization camera is described in detail in Ref.~\cite{ONUMA201469}.

The entire experimental procedure—including rotational control of the flow channel, syringe pump operation, image acquisition, and data storage—was automated to ensure repeatability and precise synchronization between rotational scanning and optical measurements.

\begin{figure*}[t]
\centering
\includegraphics[width=2.0\columnwidth]{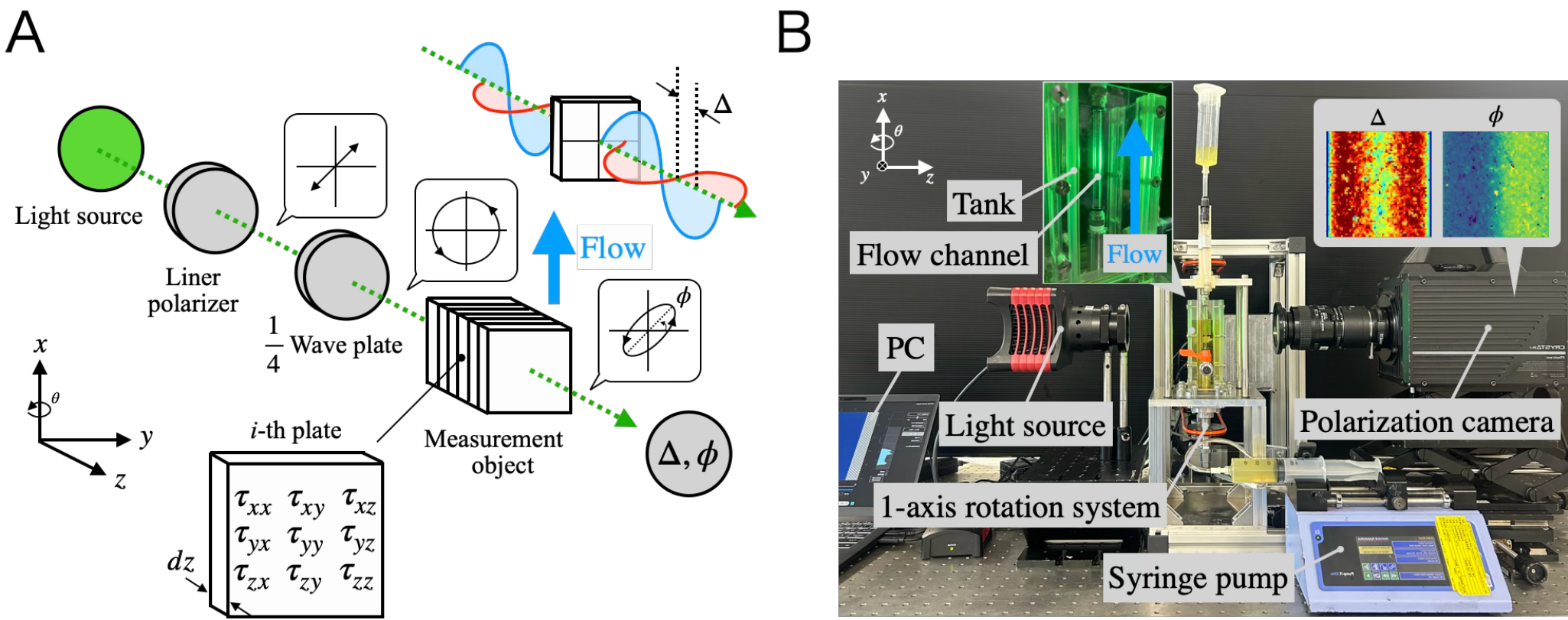}
\caption{
(A) Schematic of the optical measurement principle. 
Unpolarized light from the source is converted to circularly polarized light using a linear polarizer and a quarter-wave plate before passing through the flowing birefringent fluid. 
As the light propagates through the stressed medium, the polarization changes due to stress-induced optical anisotropy. 
The emergent light is detected by a polarization camera, and the phase retardation $\Delta$ and principal orientation $\phi$ are computed. 
The fluid domain is conceptually treated as a stack of infinitesimal slices perpendicular to the viewing direction, and the measured optical parameters represent path-integrated projections of the internal stress field.
(B) Photograph of the experimental apparatus used for rotational photoelastic tomography. 
The system consists of a circularly polarized light source, a flow channel mounted on a single-axis rotation stage, and a polarization camera. 
The motor-controlled rotation system was designed to allow controlled rotation of the liquid-filled channel while maintaining fluid sealing and optical alignment, which is
essential for acquiring projection data from multiple viewing angles.
The flow is driven by a syringe pump, while the surrounding tank contains index-matching fluid to minimize optical distortion at the tube wall. 
Inset images show representative experimentally measured retardation $\Delta$ and orientation $\phi$ fields. 
Although the raw optical data contain measurement noise, these projections provide sufficient information for robust stress reconstruction using U-FlowPET.
}
\label{fig:Experiments}
\end{figure*}

\subsection*{Dataset}
\mbox{}

To evaluate both the fundamental performance and practical applicability of the proposed method, we constructed two types of datasets: synthetic datasets with known ground-truth stress fields and experimental datasets with realistic measurement noise and uncertainties.

\subsubsection*{Synthetic Dataset}
The synthetic dataset was generated using two flow configurations: (i) a circular pipe flow with an analytical solution for baseline validation \cite{bird_transport_2002}, and (ii) a geometrically complex curved-pipe flow obtained from numerical simulation to assess applicability to non-axisymmetric three-dimensional stress fields. 
For both cases, known three-dimensional stress fields were first constructed and then converted into two-dimensional optical observations using the forward model described in the section ``Computing Phase Retardation from Stress.''

The pipe-flow case assumes a circular tube with a 4 mm inner diameter. 
For the curved pipe, we consider a 100 mm tube with the same inner diameter. 
To focus on regions where the stress field varies in three dimensions, the long upstream and downstream straight sections were excluded from the reconstruction domain; only the curved region was used for evaluation. 
Further geometric details of the curved pipe are provided in the Supplemental Material.

The properties of the working fluid were assumed to be identical to those used in the experiments. 
For the pipe-flow case, the velocity field of a power-law fluid was computed using the analytical expression

    \begin{equation}
        u(y, z) = \frac{3n+1}{n+1}  \frac{Q}{\pi R^2} \left[ 1 - \left(\frac{\sqrt{y^2+z^2}}{R}\right)^{\frac{n+1}{n}} \right],
    \end{equation}
from which the shear-rate magnitude is
    \begin{equation*}
        \dot{\gamma} = \sqrt{\left(\frac{\partial u}{\partial y}\right)^2 + \left(\frac{\partial u}{\partial z}\right)^2}.
    \end{equation*}
The corresponding stress components were calculated using the power-law constitutive relation
    \begin{equation*}
        \tau_{xy} =  K \dot{\gamma}^{n-1} \frac{\partial u}{\partial y}, \quad
        \tau_{xz} =  K \dot{\gamma}^{n-1} \frac{\partial u}{\partial z}.
    \end{equation*}
For the curved-pipe configuration, the stress field was obtained from numerical simulation using COMSOL Multiphysics by solving the steady incompressible Navier--Stokes equations. 
A mean velocity $U_{\mathrm{mean}} = Q/A$ corresponding to a flow rate of $Q=50$~mL/min was prescribed at the inlet ($x=0$). 
No-slip boundary conditions were imposed on the wall, and the outlet ($x=100$~mm) was set to atmospheric pressure.

\subsubsection*{Experimental Dataset}

Experimental datasets were obtained using the photoelastic tomography system described in the previous section in order to evaluate the robustness of the proposed method under realistic measurement noise. 
Experiments were performed in a circular pipe flow with known operating conditions.

The flow channel was rotated from $\theta = 0^\circ$ to $165^\circ$ in increments of $15^\circ$, resulting in 12 measurement angles. 
At each angle, the phase retardation and principal orientation fields were measured. 
Representative experimentally obtained retardation and orientation distributions are shown in Fig.~\ref{fig:Experiments}(B), where spatial measurement noise can be observed.

\subsection*{Implementation Details of U-FlowPET}
\mbox{}

This section describes the architecture and loss design of U-FlowPET, which enables reconstruction of three-dimensional stress fields from two-dimensional multi-angle optical measurements. 
Fig.~\ref{fig:U-FlowPET} shows a schematic overview. 
The primary optical inputs to the network are retardation $\Delta$, principal orientation $\phi$, and viewing angle $\theta$. 
Additional physical quantities, including the stress-optic coefficient $C$, flow rate $Q$, pressure drop $\Delta p$, fluid density $\rho$, and channel geometry, are used to compute the loss function. 
The channel geometry provides both boundary-condition information (e.g., no-slip walls) and the fluid-domain mask. It also determines the voxel aspect ratio of the computational domain. 
The network outputs five independent shear-stress components $\tau_{xx},\ \tau_{xy},\ \tau_{xz},\ \tau_{yy},\ \tau_{yz}$ together with the three velocity components $u,\ v,\ w$, and pressure $p$. 
The remaining stress component $\tau_{zz}$ is obtained from the trace-free condition of the extra-stress tensor for incompressible flow.

\begin{figure*}[t]
\centering
\includegraphics[width=2.0\columnwidth]{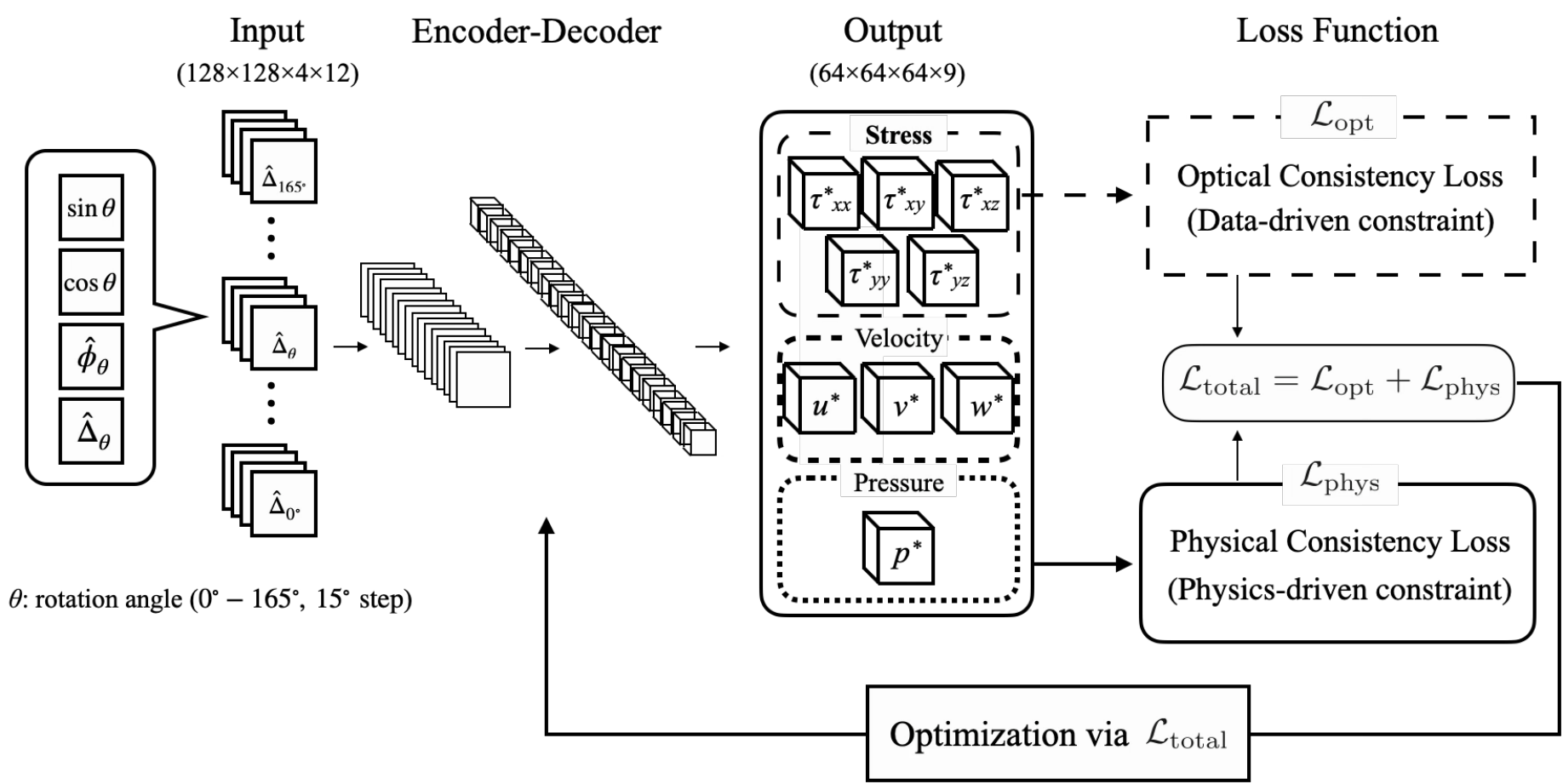}
\caption{Architecture of U-FlowPET for full-component reconstruction of three-dimensional fluid stress tensors.
Multi-angle photoelastic measurements provide two optical observables—retardation $\Delta$ and principal orientation $\phi$—as path-integrated projections of stress.
These two inputs are used to reconstruct the full three-dimensional Cauchy stress tensor field (a $3 \times 3$ tensor with six independent components), representing an intrinsically information-deficient $2 \rightarrow 6$ inverse problem. 
U-FlowPET employs a convolutional encoder–decoder network to map two-dimensional optical patterns to volumetric stress, velocity, and pressure fields. 
The reconstruction is guided by two complementary constraints: (i) an optical consistency loss enforcing consistency with measured birefringence projections and (ii) a physical consistency loss derived from the Cauchy momentum equation and continuity equation. The total loss $\mathcal{L}_{\mathrm{total}} = \mathcal{L}_{\mathrm{opt}} + \mathcal{L}_{\mathrm{phys}}$ ensures that the solution remains both data-consistent and physically admissible, enabling unsupervised, symmetry-free, and noise-robust reconstruction.}
\label{fig:U-FlowPET}
\end{figure*}

\subsubsection*{Convolutional Encoder-Decoder Architecture}

U-FlowPET employs a convolutional encoder--decoder architecture that maps multi-angle optical data to volumetric flow and stress fields. 
The encoder extracts latent features from the input tensor of size $128 \times 128 \times 4 \times 12$, which contains the measured optical observables $\Delta$ and $\phi$, together with the angular information associated with each channel rotation angle $\theta$. 
The decoder then upsamples the latent representation into a three-dimensional field of size $64 \times 64 \times 64 \times 9$, consisting of the target stress field $\boldsymbol{\tau}$ and the auxiliary fluid variables $\mathbf{v}=(u,v,w)$ and $p$ required for evaluating the physical loss. 
Further architectural details of the encoder and decoder are provided in the Supplemental Material.

To improve numerical stability and to balance the different physical quantities during training, all variables are nondimensionalized using representative scales. 
The representative pressure $P_0$, velocity $U_0$, and scale $L$ are defined as $\Delta p$, $Q / A$, and $D$, respectively, where $A$ is the cross-sectional area of the tube.
Nondimensional variables are denoted by an asterisk ($^*$).

\subsubsection*{Loss Function Design for Stress-Field Reconstruction}

U-FlowPET is trained using a hybrid loss that combines an optical consistency term, $\mathcal{L}_{\rm opt}$, with a physical consistency term, $\mathcal{L}_{\rm phys}$,
\begin{equation*}
\mathcal{L}_{\rm total} = \mathcal{L}_{\rm opt} + \mathcal{L}_{\rm phys}.
\end{equation*}
This design avoids explicit reliance on constitutive assumptions or geometric symmetry. 
Instead, $\mathcal{L}_{\rm opt}$ enforces consistency with the measured optical data, while $\mathcal{L}_{\rm phys}$ constrains the reconstructed fields to satisfy the governing equations of fluid mechanics. 
Together, these terms act as a filter that eliminates physically inadmissible solutions among the many stress fields that may be consistent with the optical projections alone.

\paragraph*{Optical consistency loss.}

The optical consistency loss is defined as
    \begin{align*}
        \mathcal{L}_{\rm opt} &= \mathcal{L}_{\Delta} + w_{\phi} \mathcal{L}_{\phi}, \\[1em]
        \mathcal{L}_{\Delta} &= \frac{1}{|\Theta|}\frac{1}{|\Omega_\theta|}  \sum_{\theta \in \Theta} \sum_{i \in \Omega_\theta} \| \hat{\Delta}_{\mathrm{pred}, \theta, i} - \hat{\Delta}_{\mathrm{in}, \theta, i} \|^2 , \\
       \mathcal{L}_{\phi} &= \frac{1}{|\Theta|}\frac{1}{|\Omega_\theta|}\sum_{\theta \in \Theta}  \sum_{i \in \Omega_\theta} \left\|\frac{1}{\pi} \mathrm{Arg} (\cos \delta \phi_{\theta, i} + j\sin\delta \phi_{\theta, i}) \right\|^2, 
    \end{align*}
where
   \begin{align*}
       \delta \phi_{\theta, i} = \phi_{\mathrm{pred}, \theta, i} - \phi_{\mathrm{in}, \theta, i},
   \end{align*}
and $\Theta = \{0^\circ, 15^\circ, 30^\circ, \dots, 165^\circ \}$ is the set of measurement angles.
            
Here, $\mathcal{L}_{\Delta}$ measures the discrepancy between the measured retardation and the retardation computed from the predicted stress field, and therefore primarily constrains stress magnitude. 
Similarly, $\mathcal{L}_{\phi}$ measures the discrepancy in principal orientation and primarily constrains stress direction. 
Retardation is normalized to the same range as the input data, $[0,1]$, before computing the error.

Because orientation is a periodic quantity, a simple subtraction cannot correctly evaluate angular error; for example, $-\pi$ and $\pi$ correspond to the same direction. 
To account for this periodicity, the angular difference is mapped to the principal interval $(-\pi,\pi]$ through the argument of its complex representation and then normalized by $\pi$. 
This formulation ensures that the optimization minimizes the shortest angular distance on the unit circle.

To reduce the influence of background noise, the optical loss is evaluated only in the foreground mask region $\Omega_{\theta}$, defined as pixels with nonzero measured retardation for each rotation angle $\theta$.

The orientation loss is additionally multiplied by a dynamic weight $w_{\phi}$. 
This is motivated by the fact that current stress--optic theory does not fully capture the alignment behavior of dispersed particles under shear \cite{mcafee1974scattered,Lane2022}, so that experimentally measured orientations may deviate from theoretical predictions. 
Accordingly, the orientation signal is used strongly at the early stage of training to guide the global search, and its influence is gradually reduced so that the final solution is determined primarily by the more reliable retardation signal and the physical constraints. 
Specifically, $w_{\phi}$ follows a cosine-decay schedule:

    \begin{equation*}
    w_{\phi}(E) = 
    \begin{cases} 
    1 & (0 \le E < E_1), \\[1em]
    \displaystyle \frac{1}{2} \left[ 1 + \cos \left( \pi \frac{E - E_1}{E_2 - E_1} \right) \right] & (E_1 \le E < E_2), \\[1em]
    0 & (E_2 \le E \le E_{\text{total}}).
    \end{cases}
    \end{equation*}
Here, $E_{\rm total} = 5000$ is the number of epochs for training, and $E_1$ and $E_2$ are set to $ 0.1 E_{\rm total}$ and $0.9 E_{\rm total}$, respectively.

\paragraph*{Physical consistency loss.}

The physical consistency loss is defined as
    \begin{equation*}
        \mathcal{L}_{\rm phys} = \mathcal{L}_{\rm PDE} + \mathcal{L}_{\rm BC} + 
        \mathcal{L}_{\rm align}
    \end{equation*}
with
    \begin{equation*}
        \mathcal{L}_{\rm PDE} = \mathcal{L}_{\rm cont} + \mathcal{L}_{\rm mom}.
    \end{equation*}
The term $\mathcal{L}_{\rm PDE}$ evaluates how closely the predicted fields satisfy the governing partial differential equations of incompressible flow at each grid point and thereby guides the solution toward fluid-mechanically admissible states. 
Importantly, no specific constitutive relation appears in these terms, so the framework does not require prior knowledge of the constitutive model.

The continuity and momentum losses are written as
    \begin{equation*}
        \mathcal{L}_{\rm cont} = \frac{1}{|\Omega_{\rm fluid}|} \sum_{i \in \Omega_{\rm fluid}} \left\| \nabla^* \cdot \mathbf{v}^*_i \right\|^2,
    \end{equation*}
    \begin{equation*}
        \mathcal{L}_{\rm mom} = \frac{1}{|\Omega_{\rm fluid}|} \sum_{i \in \Omega_{\rm fluid}} \left\| (\mathbf{v}^* \cdot \nabla^*) \mathbf{v}^* + \text{Eu} (\nabla^* p - \nabla^* \cdot \boldsymbol{\tau}^*) \right\|^2.
        \label{eq:L_mom}
    \end{equation*}
Here, $\Omega_{\rm fluid}$ denotes the set of grid points inside the fluid region, $|\Omega_{\rm fluid}|$ its cardinality, and $\text{Eu} = P_0/\rho U_0^2$ the Euler number.
$\mathcal{L}_{{\rm cont}}$ enforces incompressibility, while $\mathcal{L}_{{\rm mom}}$ enforces momentum balance through the steady Cauchy momentum equation without body force. 
While the Cauchy momentum equation is generally written in terms of the full stress tensor $\boldsymbol{\sigma}$ as in Eq. (1), the deviatoric stress tensor $\boldsymbol{\tau}$ and the pressure $p$ are used explicitly in $\mathcal{L}_{\rm mom}$ to improve numerical stability and promote stable optimization.
A key point is that stress is not computed from velocity gradients; rather, stress is treated as an independent variable that must satisfy momentum balance. 

\paragraph*{Boundary-condition loss.}

Boundary conditions are enforced through
\begin{equation*}
        \mathcal{L}_{\rm BC} = \mathcal{L}_{Q} +  \mathcal{L}_{\rm press} + \mathcal{L}_{\rm wall}.
\end{equation*}
The flow-rate loss is defined as
    \begin{equation*}
    \begin{aligned}
        \mathcal{L}_{Q} = \left( \sum_{i \in \Omega_{\mathrm{in}}} u^*_i \Delta A^*_i - Q^* \right)^2　+　\left( \sum_{i \in \Omega_{\mathrm{out}}} u^*_i \Delta A^*_i - Q^* \right)^2,
    \end{aligned}
    \end{equation*}
where $\Omega_{\mathrm{in}}$ and $\Omega_{\mathrm{out}}$ denote the inlet and outlet cross-sectional grid sets, $\Delta A^*$ is the local nondimensional area element, and $Q^* = Q/(U_0D^2)$ is the prescribed nondimensional flow rate.

The pressure-drop loss is written as
\begin{equation*}
        \mathcal{L}_{\rm press} = \left(\frac{1}{|\Omega_{\rm in}|}\sum_{i \in \Omega_{\rm in}} {p}^*_i - \Delta p^*\right)^2 + \left(\frac{1}{|\Omega_{\rm out}|}\sum_{i \in \Omega_{\rm out}}p^*_i\right)^2,
    \end{equation*}
where $p^*$ is non-dimensional pressure defined by $p^* = (p - p_{\rm out}) / P_0$, and $p_{\rm out}$ denotes the prescribed outlet pressure.

The no-slip wall loss is
    \begin{equation*}
        \mathcal{L}_{\rm wall} = \frac{1}{|\Omega_{\rm wall}|} \sum_{i \in \Omega_{\rm wall}} \|\mathbf{v}^*_i\|^2,
    \end{equation*}
where $\Omega_{\rm wall}$ denotes the set of grid points located within one voxel of the channel wall.

\paragraph*{Alignment loss.}

Finally, we introduce an alignment loss,
    \begin{equation*}
        \mathcal{L}_{\rm align} = \frac{1}{|\Omega_{\rm fluid}|} \sum_{i \in \Omega_{\rm fluid}} \frac{1}{2}\left( 1 - \frac{\boldsymbol{\tau}^*_i : \mathbf{D}^*_i}{\|\boldsymbol{\tau}^*_i\|_F \|\mathbf{D}^*_i\|_F} \right),
    \end{equation*}
where
    \begin{equation*}
        \mathbf{D}^* = \frac{1}{2} \left\{ \nabla^* \mathbf{v}^* + (\nabla^* \mathbf{v}^*)^T \right\}
    \end{equation*}
is the nondimensional strain-rate tensor, and $\| \cdot \|_F$ denotes the Frobenius norm.

This term encourages coaxiality between the deviatoric stress tensor $\boldsymbol{\tau}$ and the strain-rate tensor $\mathbf{D}$ without explicitly introducing a constitutive equation. 
It therefore provides a weak physical prior on stress orientation and helps compensate for limitations of the present stress–optic model in describing orientation-related optical responses.

\subsection*{Computational Environment and Hyperparameters}

Model implementation and training were carried out in a fixed computational environment to ensure numerical consistency and reproducibility. 
Because U-FlowPET reconstructs volumetric fields while enforcing physics-based constraints in three dimensions, the training required substantial GPU memory. 
All computations were therefore performed using an NVIDIA RTX A6000 GPU with 48~GB of VRAM.

Network optimization was performed using the Adam optimizer with a learning rate of $1 \times 10^{-4}$ and a batch size of 1. 
Training was run for 5000 epochs, over which stable convergence of the total loss was confirmed.

\subsection*{Current limitations and Future Prospects}

The present framework is developed under several simplifying assumptions. 
Specifically, the method assumes incompressible flow, steady-state conditions, and a viscous fluid without elastic effects. 
In addition, the optical model employed in this study assumes that the phase retardation does not exceed $\lambda/4$ and that the rotation of the stress principal axes along the optical path remains smaller than $\pi/6$.

Because the primary target application of the present framework is incompressible flows such as intravascular hemodynamics, compressibility effects are not considered in the current formulation. 
Nevertheless, extension to compressible flows may be possible by incorporating additional physical constraints in the loss function.

In extremely complex three-dimensional flows such as near inflow regions of cerebral aneurysms, large rotations of the stress principal axes may occur along the optical path. 
Under such conditions, the linearized stress-optic propagation model used in Eqs.~(\ref{eq:integral_stress_cos}) and (\ref{eq:integral_stress_sin}) may no longer be valid. 
A more general treatment based on nonlinear optical propagation using the Mueller matrix formalism \cite{Riera1969} would then be required.

Extension to unsteady flows would require incorporating temporal dimensions into the network architecture and accounting for particle orientation history effects in the optical response. 
Similarly, extension to viscoelastic fluids would require a new framework capable of representing elastic stresses beyond the viscous formulation considered here.
To address these multifaceted complexities, future work may explore adopting machine learning methods that satisfy symmetry and conservation laws \cite{horie2022physicsembedded, horie2024graph}.
}

\showmatmethods{} 

\showacknow{} 

\bibliography{ref_main}

\end{document}



\maketitle

\SItext

\section*{1. Sample preparation and rheological characterization}
The working fluid used in this study consisted of a suspension of cellulose nanocrystals (CNC, 0.3 wt.\%) dispersed in an aqueous sodium iodide (NaI; 57.6 wt.\%). The solution preparation followed the procedure reported by Muto et al.\cite{Muto2025}, with appropriate modifications for the present experimental conditions. 
To minimize optical refraction at the channel walls during measurements, the same NaI solution without CNC was used as a refractive index-matching fluid surrounding the test section.

For the preparation of the solutions, the CNC suspension and the NaI solution were first prepared separately and then mixed to obtain the working fluid with the desired composition. 
The NaI solution was prepared by dissolving NaI in deionized water, whereas the CNC suspension was prepared by dispersing CNC in deionized water under continuous stirring until a homogeneous suspension was obtained.


The rheological properties of the working fluid were measured using a stress-controlled rheometer (MCR302, Anton Paar) equipped with a cone-plate geometry (CP50-0.5, Anton Paar). 
The measured shear viscosity decreased with increasing shear rate, indicating typical shear-thinning behavior. 
The relationship between shear viscosity $\eta$ and shear rate $\dot{\gamma}$ was therefore modeled using the power-law equation $\eta= K \dot{\gamma}^{n-1}$.
To estimate the relevant shear-rate range for the pipe flow experiments, the apparent wall shear rate for the laminar pipe flow was approximated as $\dot{\gamma}_a = {4Q}/{\pi R^3}$, where $Q$ is the volumetric flow rate and $R$ is the radius of the tube. 
Under standard experimental conditions ($Q=50$ mL/min. and $R=2$ mm), the characteristic shear rate is approximately 130~$\text{s}^{-1}$, which was used as a reference to determine the fitting range.

\begin{figure}[ht]
\centering
\includegraphics[width=0.75\textwidth]{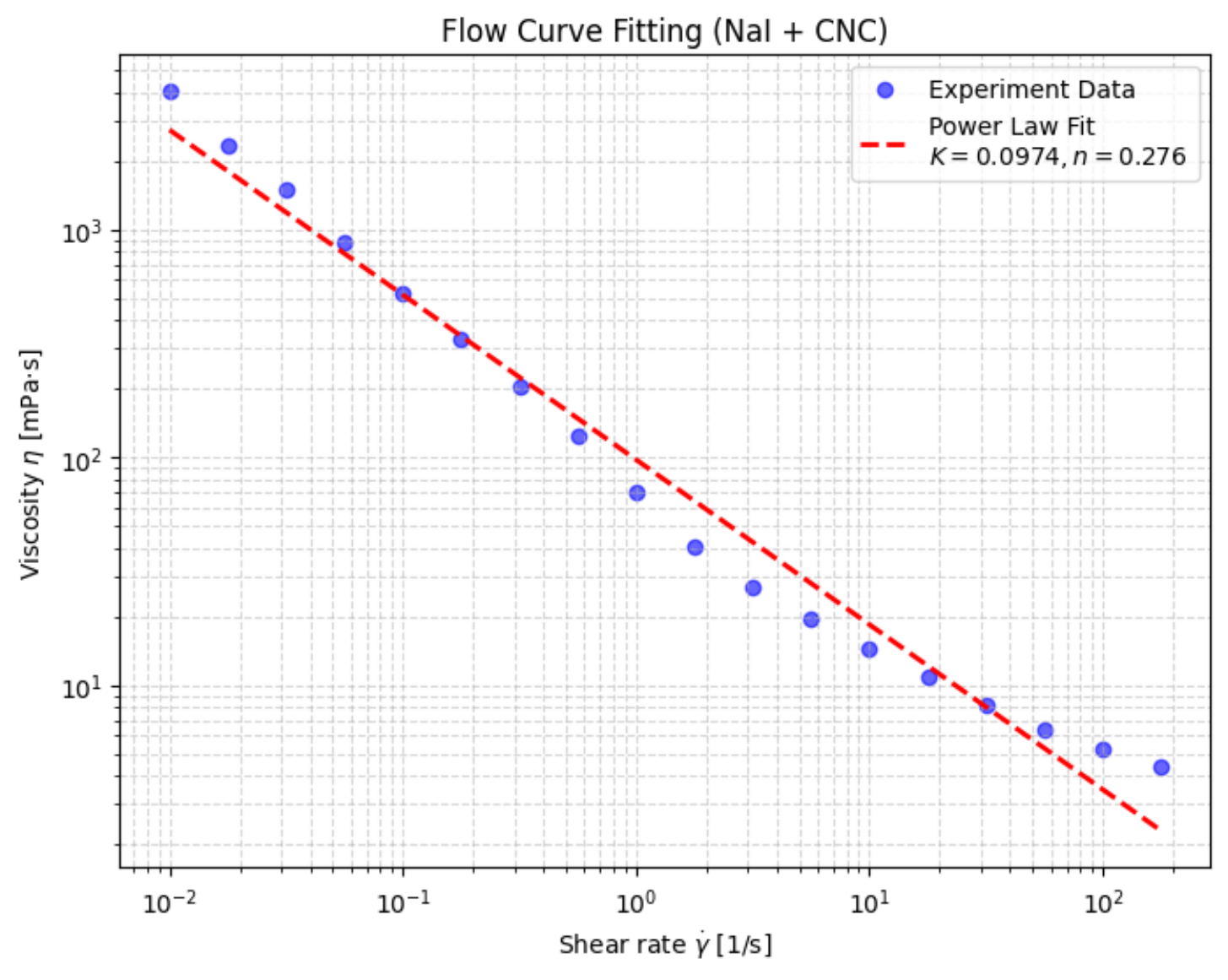}
\caption{Rheological characterization of the NaI–CNC suspension. Experimental data are shown as blue markers, while the dashed red line represents the fitted power-law model ($K = 0.0974$, $n = 0.276$). The suspension exhibits pronounced shear-thinning behavior over the measured shear-rate range.
}
\label{fig:reometer_result}
\end{figure}

The fitting procedure used rheometer data that span shear rates from 0.01 to $178\ \text{s}^{-1}$ to capture low- and high-shear regimes. 
A least-squares fit was performed in log-log space between viscosity $\eta$ and shear rate $\dot{\gamma}$. 
Figure~\ref{fig:reometer_result} shows the experimental data (blue markers) and the fitted power-law model (red dashed line), resulting in $K = 0.0974$, $n = 0.276$.
These results confirm that the working fluid exhibits shear-thinning behavior and can be reasonably described by a power-law model within the shear-rate range relevant to the present experiments.

\newpage

\section*{2. Identification of Stress-Optic Coefficients}

\subsection*{Calibration procedure}

The stress-optic coefficients were determined by fitting the theoretical retardation distribution predicted from the analytical stress field to the experimentally measured retardation profiles.

As described in the main text, the stress distribution inside the circular pipe was first computed from the analytical solution of power-law fluid flow in a cylindrical tube. 
Based on this theoretical stress field, the stress-optic coefficients $C_1$ and $C_2$ were treated as unknown parameters and determined by fitting the predicted retardation profiles to the experimentally measured data.

For the calibration experiment, the circular tube was rotated from $0^\circ$ to $180^\circ$ in increments of $15^\circ$, resulting in 13 independent viewing directions. 
At each rotation angle, the retardation field inside the fluid region was extracted and cropped into a square $N \times N$ image. 
To reduce noise and obtain a representative one-dimensional signal, the retardation field was averaged along the flow direction to produce a line profile.

The stress-optic coefficients $C_1$ and $C_2$ were then estimated by fitting the theoretical retardation distribution to the measured line profiles. 
To account for photon shot noise inherent in optical measurements, the fitting was performed using a nonlinear least-squares approach with Poisson weighting. 
The fitting involved minimizing the chi-squared error objective function:

\[
\chi^2 = \sum_{i=1}^{N} w_i (\Delta_{\mathrm{expt},i} - \Delta_{\mathrm{fit},i})^2,
\]

where the Poisson weight is defined as

\[
w_i = \frac{1}{\Delta_{\mathrm{expt},i}}.
\]

This weighting scheme allows relative errors in both high- and low-retardation regions to be evaluated more appropriately than the standard least-squares fitting.

The fitting procedure was performed independently for all 13 viewing angles, yielding a set of $C_1$ and $C_2$ values for each orientation. 
The final stress-optic coefficients were obtained by averaging the fitted values over all viewing angles.

Figure~\ref{fig:C1,C2fitting} shows the processed experimental retardation profiles together with the theoretical curves calculated using the fitted coefficients $C_1 = 1.32 \times 10^{-5}$ and $C_2 = 5.13 \times 10^{-5}$. 
The close agreement between the measured and predicted distributions confirms the validity of the identified stress-optic coefficients.

\begin{figure}[ht]
\centering
\includegraphics[width=0.75\textwidth]{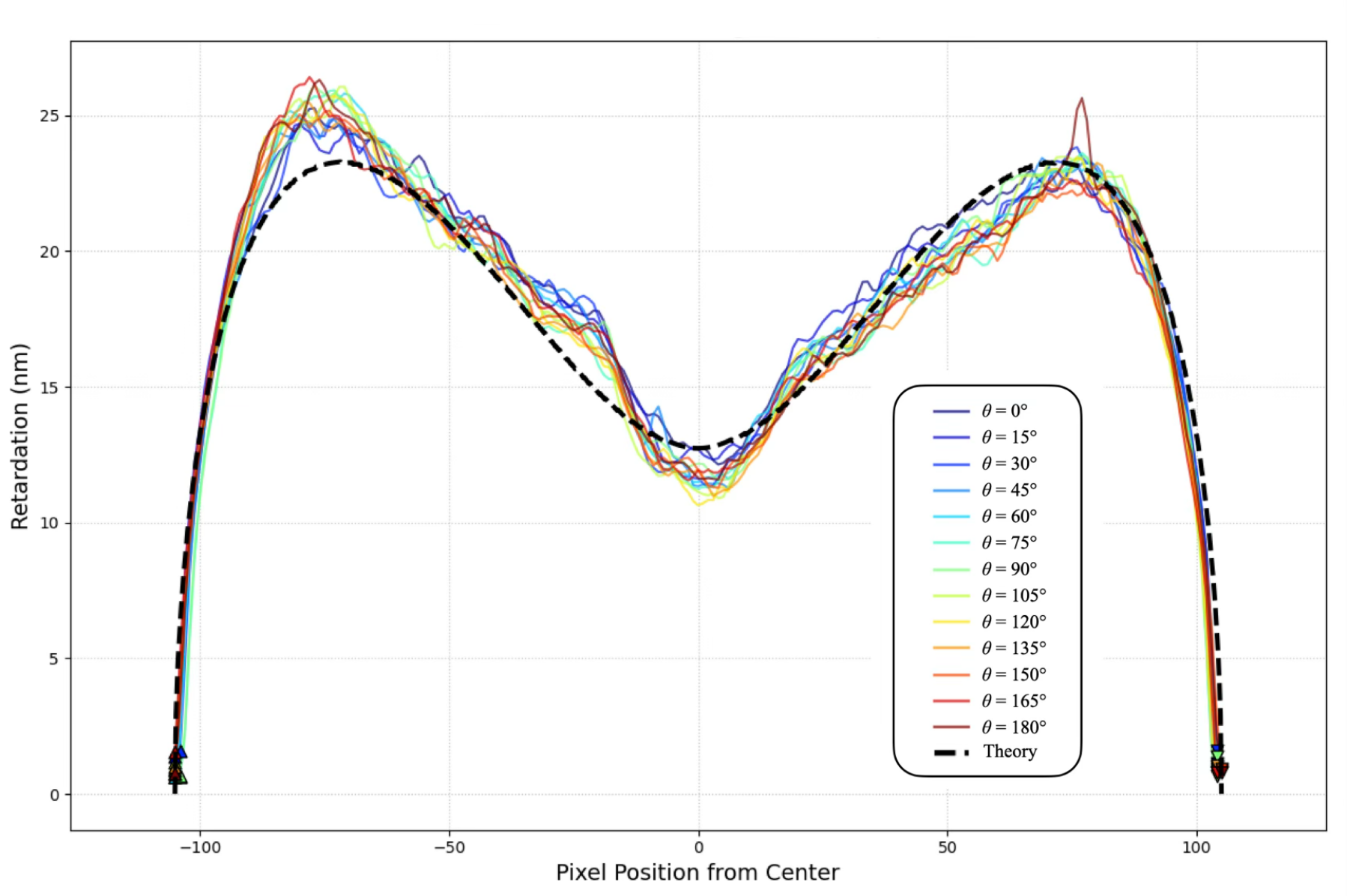}
\caption{Calibration of the stress-optic coefficients. Colored curves show the measured retardation profiles obtained from 13 viewing angles, while the dashed black line indicates the theoretical retardation distribution computed using the fitted coefficients $C_1 = 1.32 \times 10^{-5}$ and $C_2 = 5.13 \times 10^{-5}$.}
\label{fig:C1,C2fitting}
\end{figure}

\newpage

\section*{3. Geometry of the Curved Pipe Used to Generate Synthetic Data}

\begin{figure}[ht]
\centering
\includegraphics[width=0.8\textwidth]{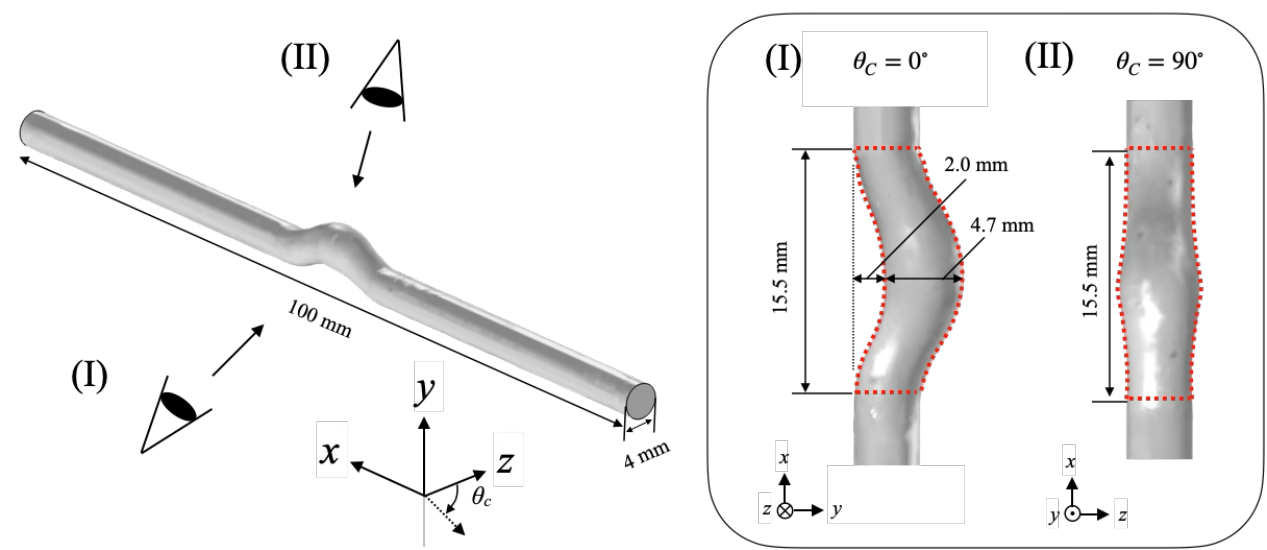}
\caption{Geometry of the curved pipe used to generate synthetic datasets. 
The pipe has an inner diameter of 4~mm and a total length of 100~mm. 
Two views are shown corresponding to coordinate rotations $\theta_c=0^\circ$ (I) and $\theta_c=90^\circ$ (II). 
The red dashed region indicates the reconstruction domain near the curved section where the stress field exhibits strong 3D variation.}
\label{fig:cos_curve_ryuro}
\end{figure}

Fig.~\ref{fig:cos_curve_ryuro} shows the geometric configuration of the curved pipe used to generate the synthetic dataset. 
The pipe has a circular cross section with an inner diameter of 4~mm and a total length of 100~mm.

The geometry of the curved section is illustrated from two viewing directions obtained by rotating the coordinate system about the $x$ axis through an angle $\theta_c$. 
Panel (I) corresponds to $\theta_c = 0^\circ$, while panel (II) corresponds to $\theta_c = 90^\circ$. 
These views highlight the three-dimensional (3D) curvature of the channel.

The red dashed region indicates the reconstruction domain used in this study. 
This region corresponds to the vicinity of the curved section, where the stress field exhibits strong 3D variations. 
To reduce computational cost while preserving the essential flow physics, the long, straight upstream and downstream segments were excluded from the reconstruction domain, and computational resources were concentrated in the curved region.

\section*{4. Detailed Architecture of the U-FlowPET Network}

\begin{figure}[ht]
\centering
\includegraphics[width=0.75\textwidth]{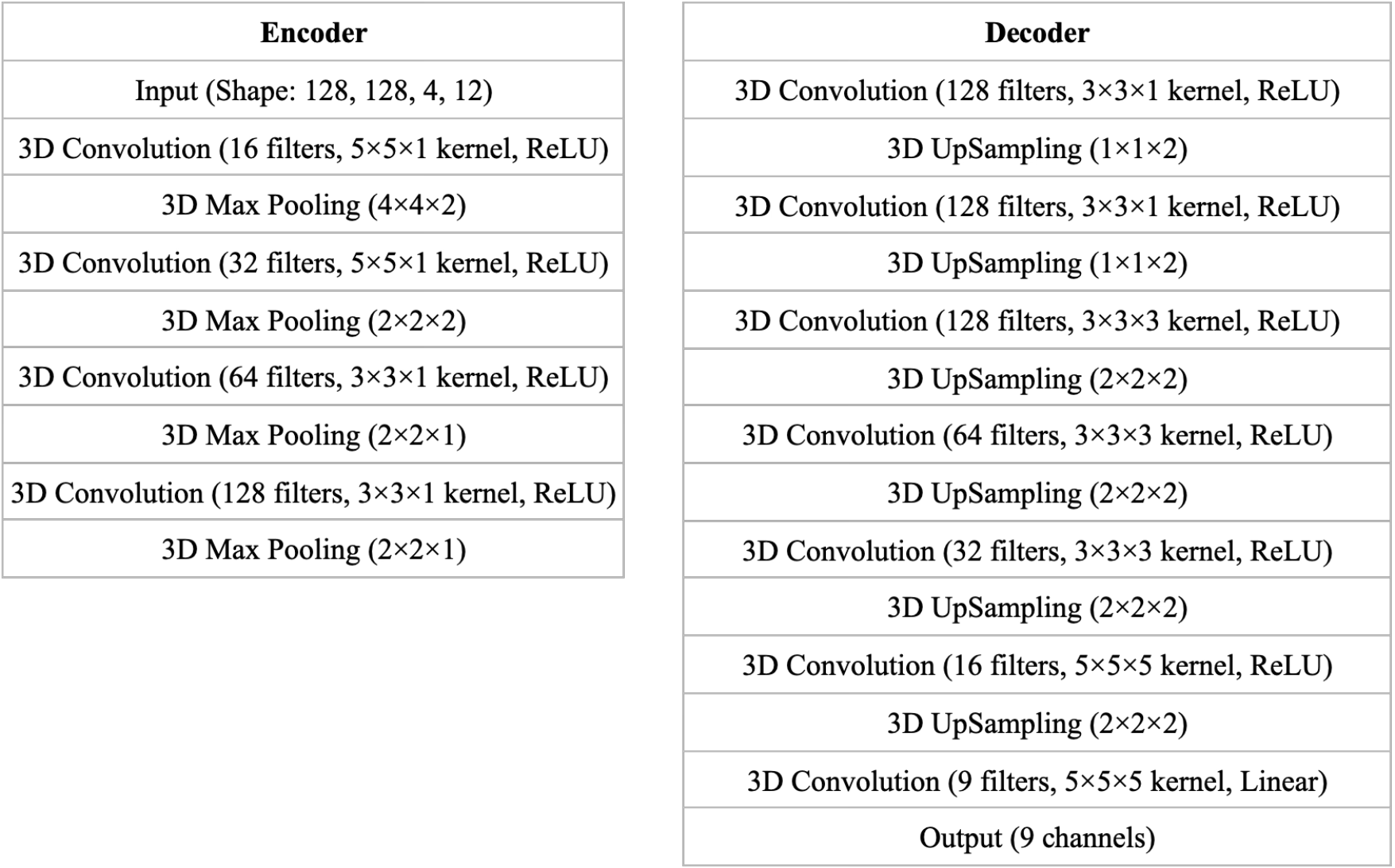}
\caption{Detailed architecture of the 3D CNN used in U-FlowPET. 
The encoder extracts features from multi-angle optical inputs, while the decoder reconstructs volumetric stress, velocity, and pressure fields.}
\label{fig:3DCNN_based_model}
\end{figure}

The proposed U-FlowPET model is based on a 3D convolutional neural network (3D CNN). 
The detailed network architecture is illustrated in Fig.~\ref{fig:3DCNN_based_model}.

The input layer has dimensions $128 \times 128 \times 4 \times 12$, representing multi-angle optical observations. 
For each of the 12 viewing directions, the input channels consist of the measured retardation $\Delta$, principal orientation $\phi$, and the angular information of the flow channel expressed as $\sin\theta$ and $\cos\theta$. 
These inputs provide both the optical observations and the geometric information required for reconstruction.

The encoder extracts hierarchical spatial features from the input tensor using a sequence of 3D convolution and max-pooling layers. 
To emphasize spatial patterns within each angular slice, the convolution kernels in the encoder are designed with unit depth in the angular dimension. 
This configuration allows the network to prioritize the extraction of spatial optical features within each view before integrating information across different viewing angles.

The decoder reconstructs volumetric fields from the latent representation through a sequence of 3D upsampling and convolution layers. 
Unlike the encoder, the decoder employs full 3D convolution kernels, enabling the network to reconstruct spatially continuous 3D fields.

The output layer contains nine channels corresponding to the reconstructed physical variables. 
These include the five independent components of the nondimensional extra-stress tensor, the three components of the velocity vector $(u,v,w)$, and the pressure field $p$.

Rectified Linear Unit (ReLU) activation functions were used in all hidden layers, while a linear activation function was applied to the output layer to avoid restricting the range of physical quantities. 
To preserve spatial resolution throughout the network, the same padding was applied in all convolution layers.

\FloatBarrier

\movie{\textbf{Multi-angle photoelastic observables ($\Delta$ and $\phi$) for U-FlowPET reconstruction.} 
This movie displays the path-integrated optical projections of a flowing birefringent fluid captured across multiple viewing angles. 
The left panel shows the retardation $\Delta$, representing the magnitude of the optical phase shift, while the right panel shows the principal orientation $\phi$, indicating the projected stress direction. 
These multi-angle data serve as raw input for the U-FlowPET framework to reconstruct the 3D Cauchy stress tensor field. 
The projection sequence covers a rotation range from 0$^\circ$ to 165$^\circ$ with an angular step of 15$^\circ$.}





\bibliography{ref_SI}